\documentstyle[aps,floats,graphicx,tighten]{revtex}

\begin{document}

\def\nuc#1#2{${}^{#1}$#2}
\def\aprgt{\buildrel > \over {_{\sim}}}

\draft

\twocolumn[\hsize\textwidth\columnwidth\hsize\csname@twocolumnfalse\endcsname

\begin{flushright}
{\footnotesize astro-ph/9907131}
\end{flushright}

\title{Measurement of the Solar Neutrino Capture Rate by SAGE and
       Implications \\ for Neutrino Oscillations in Vacuum}

\author{J.\,N.\,Abdurashitov,$^1$ T.\,J.\,Bowles,$^2$ M.\,L.\,Cherry,$^3$
        B.\,T.\,Cleveland,$^4$ R.\,Davis, Jr.,$^4$ S.\,R.\,Elliott,$^5$
        V.\,N.\,Gavrin,$^1$ S.\,V.\,Girin,$^1$ V.\,V.\,Gorbachev,$^1$
        T.\,V.\,Ibragimova,$^1$ A.\,V.\,Kalikhov,$^1$ N.\,G.\,Khairnasov,$^1$
        T.\,V.\,Knodel,$^1$ K.\,Lande,$^4$ I.\,N.\,Mirmov,$^1$
        J.\,S.\,Nico,$^6$ A.\,A.\,Shikhin,$^1$ W.\,A.\,Teasdale,$^2$
        E.\,P.\,Veretenkin,$^1$ V.\,M.\,Vermul,$^1$
        D.\,L.\,Wark,$^{2,}$\cite{Wark} P.\,S.\,Wildenhain,$^4$
        J.\,F.\,Wilkerson,$^5$ V.\,E.\,Yants,$^1$ and G.\,T.\,Zatsepin$^1$ \\
        (SAGE Collaboration)}

\address{$^1$ Institute for Nuclear Research, Russian Academy of Sciences,
              117312 Moscow, Russia}
\address{$^2$ Los Alamos National Laboratory, Los Alamos, New Mexico 87545}
\address{$^3$ Louisiana State University, Baton Rouge, Louisiana 70803}
\address{$^4$ University of Pennsylvania, Philadelphia, Pennsylvania 19104}
\address{$^5$ University of Washington, Seattle, Washington 98195}
\address{$^6$ National Institute of Standards and Technology, Gaithersburg,
              Maryland 20899}

\date{Received 12 July 1999}

\maketitle

\begin{abstract}
       The Russian-American solar neutrino experiment has measured the
       capture rate of neutrinos on metallic gallium in a
       radiochemical experiment at the Baksan Neutrino Observatory.
       Eight years of measurement give the result $67.2 ^{+7.2+3.5}
       _{-7.0-3.0}$ solar neutrino units, where the uncertainties are
       statistical and systematic, respectively.  The restrictions
       these results impose on vacuum neutrino oscillation parameters
       are given.
\end{abstract}


] 

Although standard solar models (SSM) based on nuclear fusion have had great
success in explaining many observed properties of the Sun, their prediction
of the solar neutrino flux is not consistent with experimental measurements.
The Homestake chlorine experiment~\cite{CLE98}, the water Cherenkov detectors
Kamiokande~\cite{FUK96} and Super-Kamiokande~\cite{FUK981}, and the Ga
experiments SAGE~\cite{ABD94,ABD96,ABD99}, and GALLEX~\cite{HAM99} have all
measured a neutrino detection rate considerably below SSM predictions.  In
view of the recent very strong evidence for oscillations of atmospheric
neutrinos \cite{FUK982}, it seems reasonable to suppose that the deficit of
solar neutrinos may also be the result of neutrino oscillations.

In this Letter we present results of the ongoing SAGE experiment and consider
its implications on the widely discussed hypothesis of vacuum oscillations.

Ga experiments detect neutrinos by the reaction
\nuc{71}{Ga}$(\nu_e,e^-)$\nuc{71}{Ge}.  They are the only presently operating
experiments with a sufficiently low threshold (233 keV) to be able to measure
the low-energy neutrinos from proton-proton ($pp$) fusion -- the major energy
producing reaction in the Sun.  SSM calculations~\cite{BAH98,TUR98} predict
that the total expected capture rate in \nuc{71}{Ga} is 129 solar neutrino
units (SNU), of which 69.6 SNU arise from the $pp$ neutrinos, with
significant contributions from the \nuc{7}{Be} and \nuc{8}{B} neutrinos (34.4
SNU and 12.4 SNU, respectively), and lesser contributions from the CNO and
$pep$ neutrinos (9.8 SNU and 2.8 SNU, respectively).  [1 SNU = (10$^{-36}$
interactions/(s)/target atom).]

\begin{table}[t]
\squeezetable
\caption{Results for monthly and bimonthly combinations of SAGE data.  Runs
are assigned to each time interval by their mean exposure time.  The $1/R^2$
dependence due to the Earth-Sun distance variation has been removed from the
capture rate.}
\label{monthly}
\begin{tabular}{c d d r @{--} d}
Exposure  & Number of & \multicolumn{3}{c}{Capture rate (SNU)} \\ \cline{3-5}
interval  & data sets & Best fit & \multicolumn{2}{c}{68\% conf.~range}  \\
\hline
Jan     &   7. &   47. &  24 &  74. \\
Feb     &   6. &   41. &  20 &  63. \\
Mar     &   3. &  198. & 137 & 266. \\
Apr     &   5. &   41. &  22 &  63. \\
May     &   6. &   83. &  58 & 111. \\
Jun     &   3. &   37. &   3 &  80. \\
Jul     &   9. &   40. &  22 &  62. \\
Aug     &   9. &   79. &  57 & 102. \\
Sep     &  12. &   63. &  47 &  82. \\
Oct     &  11. &   64. &  42 &  90. \\
Nov     &   9. &   73. &  52 &  96. \\
Dec     &   8. &  123. &  95 & 153. \\
\\
Jan+Feb &  13. &   44. &  28 &  60. \\
Mar+Apr &   8. &   70. &  48 &  94. \\
May+Jun &   9. &   71. &  50 &  95. \\
Jul+Aug &  18. &   60. &  45 &  77. \\
Sep+Oct &  23. &   64. &  50 &  79. \\
Nov+Dec &  17. &   95. &  77 & 113. \\
\\
Feb+Mar &   9. &   69. &  48 &  92. \\
Apr+May &  11. &   60. &  44 &  78. \\
Jun+Jul &  12. &   39. &  23 &  59. \\
Aug+Sep &  21. &   70. &  57 &  84. \\
Oct+Nov &  20. &   69. &  54 &  86. \\
Dec+Jan &  15. &   88. &  70 & 106.
\end{tabular}
\normalsize
\end{table}

\begin{figure*}
\begin{center}
\includegraphics[width=5.000in]{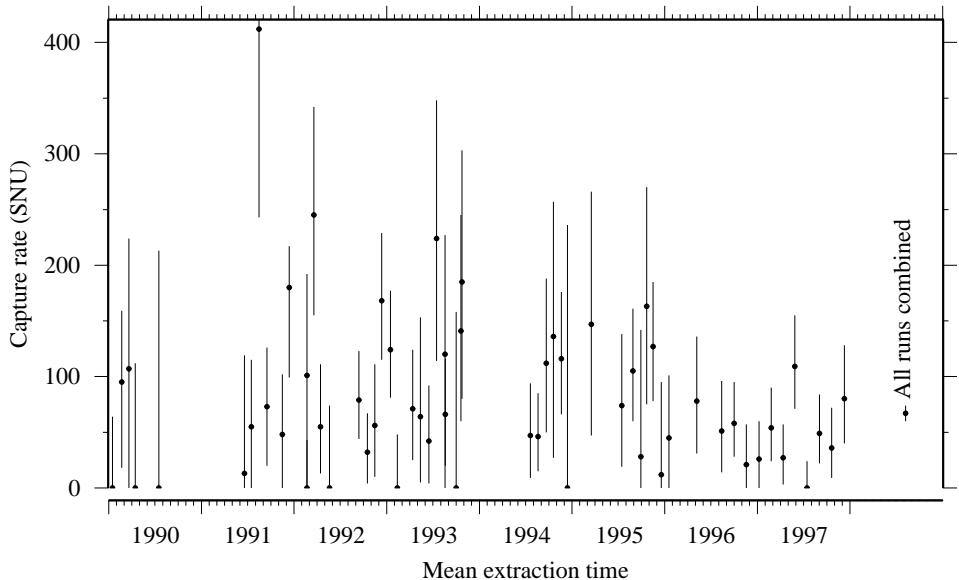}
\end{center}
\caption{Capture rate for each extraction as a function of time.  All error
bars represent statistical uncertainties only.}
\label{All_extr}
\end{figure*}

A detailed discussion of the SAGE experimental procedures, including the
chemical extraction, low-background counting of \nuc{71}{Ge}, data analysis
methods, and systematic effects, is given in~\cite{ABD99}.  The combined
result from 88 separate counting data sets is $67.2 ^{+7.2 +3.5} _{-7.0
-3.0}$ SNU.  The dominant contributions to the systematic uncertainty come
from the Ge extraction efficiency and the \nuc{71}{Ge} counting efficiency.
The individual measurement results are plotted in Fig.~\ref{All_extr}.

The SAGE result of 67.2 SNU is approximately $7 \sigma$ lower than SSM
predictions.  It is almost impossible to reconcile this discrepancy by an
alteration of the astrophysical components of the SSM.  If one artificially
sets the rate of the \nuc{3}{He}$(\alpha,\gamma)$\nuc{7}{Be} reaction to
zero, so that the \nuc{7}{Be} and \nuc{8}{B} neutrinos are eliminated, then
solar models predict~\cite{BAH97} that the Ga experiment should measure $88.1
^{+3.2}_{-2.4}$ SNU, more than $2 \sigma$ greater than our result.  If, in
addition, all the cross sections for the CNO reactions are set to zero, so
that the Sun produces only $pp$ and $pep$ neutrinos, then the Ga experiment
should measure $79.5 ^{+2.3}_{-2.0}$ SNU, about $1.5 \sigma$ above our
result.  Since the $pp$ rate is well determined by the solar luminosity, the
deficit of solar neutrinos observed in the Ga experiment implies that new
physics beyond the standard model of the electroweak interaction is required
to understand the solar neutrino spectrum.

A credible explanation of the solar neutrino problem that does not contradict
any other known phenomena is to assume that the neutrinos produced in the Sun
have changed flavor by the time they reach the Earth.  There are several ways
in which such neutrino oscillations may occur.  In one type,
Mikheyev-Smirnov-Wolfenstein (MSW) oscillations, the solar $\nu_e$ transforms
into other flavor neutrinos or a sterile neutrino as it passes through a thin
resonance region near the solar core.  In the second type, vacuum
oscillations, the neutrino changes flavor in the vacuum between the Sun and
the Earth.  In another type, resonant spin flavor conversion, the electron
neutrino, provided it has a suitably large magnetic moment, transforms into
other species undetectable in Ga as it passes through the solar magnetic
field.

\begin{figure}[t]
\begin{center}
\includegraphics[width=3.375in]{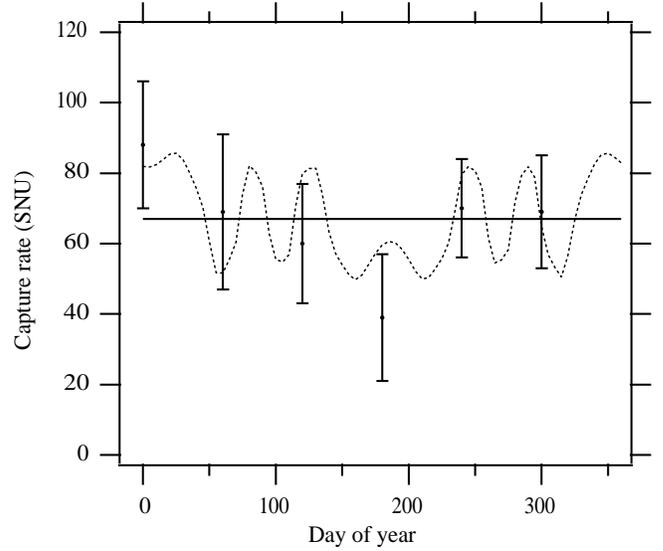}
\end{center}
\caption{Solar neutrino detection rate vs time of year for the February+March
grouping in Table~\ref{monthly}.  Superimposed are theoretical curves for a
constant capture rate of 67 SNU (solid line) and a vacuum oscillation
solution with $\Delta m^2 = 1.2 \times 10^{-9}$ eV$^2$ and $\sin^2 2\theta =
0.94$ (dashed line).}
\label{VO}
\end{figure}

Oscillations between two neutrino species are characterized by two
parameters: $\Delta m^2$, the difference of the eigenstate masses, and
$\theta$, the mixing angle between the mass eigenstates.  The Ga experiments,
sensitive to the low-energy $pp$ and \nuc{7}{Be} neutrinos, combined with the
high-energy response of the Cl and Super-Kamiokande experiments,
substantially restrict the allowed range of $\Delta m^2$ and $\theta$ for all
oscillation scenarios.  The regions of parameter space that are consistent
with all solar neutrino experiments have been well discussed in the
literature -- see~\cite{BIL98} for a comprehensive review and references to
original papers.  At the present time there is no evidence that favors any
one of the various oscillation solutions over the others.

As an example of neutrino oscillations, we consider in the following the case
of vacuum oscillations (VO).  Under the VO assumption, a reasonably good fit
to the results of all solar neutrino experiments is obtained for $\Delta m^2
\simeq 6.5 \times 10^{-11}$ eV$^2$ and $\sin^2 2\theta \simeq 0.75$
\cite{BKS98}.  One predicted consequence of neutrino oscillations for
parameters in this range is a seasonal variation in the solar neutrino flux.
If such a seasonal variation were observed it would distinguish clearly
between the MSW and VO solutions as parameters in most of the MSW range give
no detectable time variation in the Ga experiment beyond that expected from
the eccentricity of the Earth's orbit.

To explore this possibility, we give in Table~\ref{monthly} the results of
the combined analysis of subsets of SAGE data that are grouped by the time of
year in which the exposure occurred.  The bimonthly grouping that combines
February and March is shown in Fig.~\ref{VO}.  This choice is arbitrary and
the qualitative conclusions we draw below are insensitive to it.
Approximating the asymmetric statistical uncertainty by a symmetric error, an
expedient analysis technique that makes details of the fit easy to elucidate
and extends readily to the analysis discussed below, these results fit quite
well ($\chi^2 = 4.9$ with 5 degrees of freedom) to a constant value of 67.2
SNU, the global best fit to the SAGE data.

Since the fit to a constant rate is quite good, there is no need to invoke VO
to explain the time dependence of the data.  Nonetheless, to see how the
neutrino parameter space is constrained by the SAGE time-of-year results, we
will fit them to the VO hypothesis.  The survival probability $P_{\nu_e
\rightarrow \nu_e}$ of an electron neutrino of energy $E$ which undergoes
vacuum oscillations can be written~\cite{Pontecorvo}

\begin{displaymath}
P_{\nu_e \rightarrow \nu_e} = 1 - \sin^2 2\theta \sin^2 (\pi R/L),
\end{displaymath}

\noindent where $R$ is the distance between the neutrino emission point and
the detector and $L$ is the neutrino oscillation length, given by $L = 2.47
E/\Delta m^2$, with $E$ in MeV, $\Delta m^2$ in eV$^2$, and $L$ in m.  Since
perihelion of the Earth's orbit occurs during the first week of January, the
Earth-Sun distance $R$ can be approximated by

\begin{displaymath}
R = 1.496 \times 10^{11}
    \left[1.0 - 0.0167 \cos\frac{2\pi(t-3.5)}{365}\right] \text{ m},
\end{displaymath}

\noindent where $t$ is the day of year.  Combining these equations leads to

\begin{eqnarray}
& & P_{\nu_e \rightarrow \nu_e}(\Delta m^2,\theta,E,t)
    = 1 - \sin^2 2\theta  \nonumber \\
& & \times \sin^2\!\left[1.90 \times 10^{11} \frac{\Delta m^2}{E}
                 \!\left(1.0-0.0167 \cos \frac{2\pi(t-3.5)}{365}
                                               \!\right)\right]\!. \nonumber
\end{eqnarray}

\noindent For $\Delta m^2 \simeq 10^{-10}$ eV$^2$ and $E \simeq 1$ MeV, the
3\% change in the Earth-Sun distance during the year can change the phase of
the term in square brackets by $\pi$.  For the \nuc{7}{Be} and $pep$ neutrino
lines, this can lead to a dramatic variation in the survival probability as
$P_{\nu_e\rightarrow\nu_e}$ varies from 1 to $1 - \sin^2 2\theta \simeq 0$.

Using this survival probability, the cross section $\sigma(E)$ for inverse
beta decay on \nuc{71}{Ga}~\cite{BAH97}, the flux~\cite{BAH98}, and the
spectral shape $F(E)$~\cite{BAH89}, the capture rate $C$ observed in the Ga
detector is given by

\begin{displaymath}
C(\Delta m^2,\theta,t) = \int {P_{\nu_e\rightarrow\nu_e}(\Delta
m^2,\theta,E,t) \sigma(E) F(E) dE}. \nonumber
\end{displaymath}

Since the $pp$, \nuc{8}{B}, and CNO neutrino sources are not lines, the
integration of their survival probability over energy gives a nearly constant
contribution that is reduced from the no-oscillation value by the factor $ 1
- \frac{1}{2} \sin^2 2\theta$.  Thus, VO can cause an overall decrease in $C$
with respect to the SSM.  Further, because of the large contribution of
\nuc{7}{Be} neutrinos to the response of the Ga detector, the rate can depend
strongly on the time of year for certain values of the oscillation
parameters.

\begin{figure}[t]
\begin{center}
\includegraphics[width=3.375in]{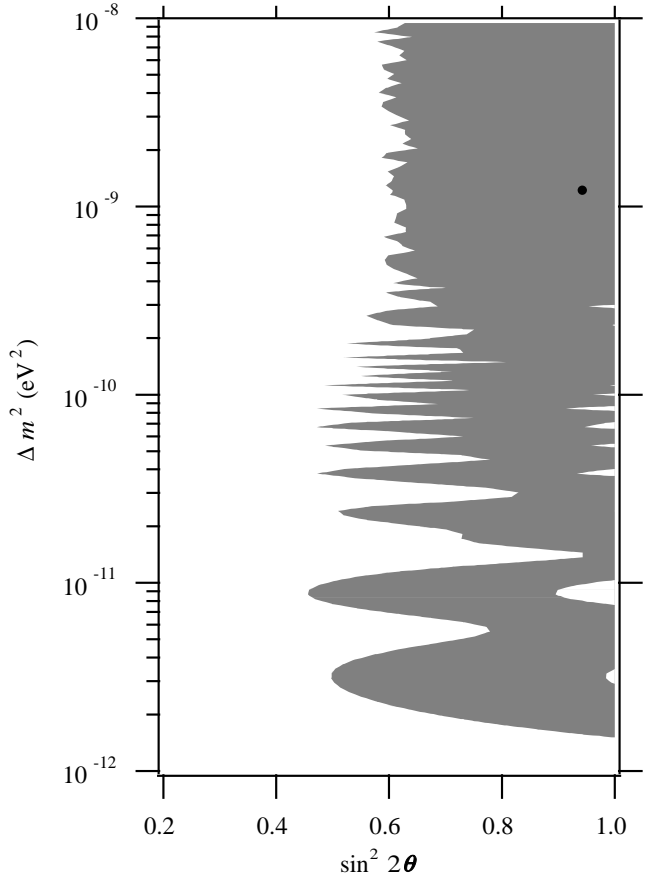}
\end{center}
\caption{Shaded area indicates the allowed region of neutrino parameters at
90\% confidence level determined from the February+March data grouping
assuming vacuum oscillations.  Black circle marks the best fit point.}
\label{allowed}
\end{figure}

To constrain the range of allowed neutrino oscillation parameters, we average
$C$ over the two-month measurement period and calculate the sum of $\chi^2$
for the 6 data points in Fig.~\ref{VO}.  The systematic uncertainty of
$\simeq 5\%$ is neglected as it is negligible compared to the $\simeq 33\%$
statistical uncertainty of each bimonthly measurement.  The fluxes predicted
by the SSM are uncertain by $\simeq 5\%$, and we ignore that uncertainty
also.  A plot of contours of $\Delta \chi^2 = 7.8$, which defines the region
of 90\% confidence for 4 degrees of freedom, is shown in Fig.~\ref{allowed}.
The overall minimum is at $\Delta m^2 = 1.2 \times 10^{-9}$ eV$^2$ and
$\sin^2 2\theta = 0.94$ and has $\chi^2 = 0.5$.  The time dependence
predicted with these parameters is shown in Fig.~\ref{VO}.  Since the $1/R^2$
dependence of the flux has been removed in the reported rates, the variation
here is solely due to vacuum oscillations.  No particular significance should
be attached to this best fit point, however, nor should our results be
interpreted as favoring any particular region in the VO allowed space.  This
is because the location of the best fit point changes depending on the way in
which runs are grouped in the time average and there are many other points in
the parameter space where the fit quality is nearly as good as at the best
fit point.  Further, for $\Delta m^2 \aprgt 5 \times 10^{-10}$ eV$^2$,
because the oscillations are so rapid, the allowed region shown in
Fig.~\ref{allowed} is determined mainly by the total observed capture rate,
with minor changes to the boundary from the time dependence.  We also need to
note that since the neutrino fluxes predicted by the SSM were used in this
analysis, these results are not model independent.

The best fit to the neutrino energy spectrum measured at Super-Kamiokande,
assuming the reduction in flux compared to the SSM is due to VO, is at
$\Delta m^2 \simeq 4.3 \times 10^{-10}$ eV$^2$ and $\sin^2 2\theta \simeq
0.87$ \cite{Smy}.  As is evident from Fig.~\ref{allowed}, this region of
neutrino parameters is compatible with the SAGE measurements.  Further
running of SAGE will reduce the uncertainties in a two month bin to about
$\pm 15$ SNU, thus restricting the total region of allowed VO parameter space
to approximately 70\% of current limits.  A further improvement of the limits
will occur by combining the measurements of both Ga experiments and
additional restriction is to be expected from much higher rate experiments
such as Super-Kamiokande, SNO, and Borexino.

In summary, the combined analysis of all experiments strongly indicates that
the solar neutrino deficit has a particle physics explanation and is a
consequence of neutrino mass.  The present experiments are, however, not yet
able to establish definitively the oscillation scenario.  Reduction of the
uncertainties of the existing experiments, and new experiments, particularly
those with sensitivity to low-energy neutrinos or to neutrino flavor, are
urgently needed.  SAGE is currently making regular solar neutrino extractions
every six weeks with $\simeq 50$ t of Ga and plans to continue these
measurements until 2006.  This will further reduce the statistical and
systematic uncertainties, thus providing greater sensitivity to the
model-independent astrophysical limit of 79.5 SNU in the Ga experiment and
further limiting possible oscillation solutions to the solar neutrino
problem.

We thank J. N. Bahcall, V. S. Berezinsky, M. Baldo-Ceolin, P. Barnes, G. T.
Garvey, W. Haxton,  V. A. Kuzmin, V. A. Matveev, L. B. Okun, V. A. Rubakov,
R. G. H. Robertson, and A. N. Tavkhelidze for their continued interest and
for fruitful and stimulating discussions.  We acknowledge the support of the
Russian Academy of Sciences, the Institute for Nuclear Research of the
Russian Academy of Sciences, the Ministry of Science and Technology of the
Russian Federation, the Russian Foundation of Fundamental Research under
Grant No.\ 96-02-18399, the Division of Nuclear Physics of the U.S.
Department of Energy, the U.S. National Science Foundation, and the U.S.
Civilian Research and Development Foundation under Award No.\ RP2-159.  This
research was made possible in part by Grant No.\ M7F000 from the
International Science Foundation and Grant No.\ M7F300 from the International
Science Foundation and the Russian Government.

\end{document}